\documentclass[twocolumn]{webofc}
\usepackage[varg]{txfonts} 
\usepackage{graphicx}
\usepackage{amsmath}
\usepackage{amssymb}

\begin{document}

\title {FRAM telescopes and their measurements of aerosol content at the Pierre Auger Observatory and at future sites of the Cherenkov Telescope Array}

\author{\firstname{Petr} \lastname{Jane\v{c}ek}\inst{1}\fnsep\thanks{\email{janecekp@fzu.cz}} for the CTA Consortium\thanks{ http://www.cta-observatory.org/} \and
\firstname{Jan} \lastname{Ebr}\inst{1} \and
\firstname{Jakub} \lastname{Jury\v{s}ek}\inst{1} \and
\firstname{Michael} \lastname{Prouza}\inst{1} \and
\firstname{Ji\v{r}\'{i}} \lastname{ Bla\v{z}ek}\inst{1} \and
\firstname{Petr} \lastname{ Tr\'{a}vn\'{i}\v{c}ek}\inst{1} \and
\firstname{Du\v{s}an} \lastname{Mand\'{a}t}\inst{1} \and
\firstname{Miroslav} \lastname{Pech}\inst{1}\ 
for the Pierre Auger Collaboration\thanks{Full author list: http://www.auger.org/archive/authors\_2018\_09.html} and the CTA Consortium\footnotemark[2]
\and 
\firstname{Sergey} \lastname{Karpov}\inst{1} \and
\firstname{Ronan} \lastname{Cunniffe}\inst{1} \and
\firstname{Martin} \lastname{Ma\v{s}ek}\inst{1} \and
\firstname{Martin} \lastname{Jel\'{i}nek}\inst{2} \and
\firstname{Ivana} \lastname{Ebrov\'{a}}\inst{1,3}
}

\institute{Institute of Physics of the Czech Academy of Sciences, Prague 
\and
Astronomical Institute of the Czech Academy of Sciences, Ond\v{r}ejov
\and
Nicolaus Copernicus Astronomical Center, Polish Academy of Sciences, Warsaw
}

\abstract{
A FRAM (F/(Ph)otometric Robotic Atmospheric Monitor) telescope is a system of a robotic mount, a large-format CCD camera and a fast telephoto lens that can be used for atmospheric monitoring at any site when information about the atmospheric transparency is required with high spatial or temporal resolution and where continuous use of laser-based methods for this purpose would interfere with other observations. The original FRAM has been operated at the Pierre Auger Observatory in Argentina for more than a decade, while three more FRAMs are foreseen to be used by the Cherenkov Telescope Array (CTA). The CTA FRAMs are being deployed ahead of time to characterize the properties of the sites prior to the operation of the CTA telescopes; one FRAM has been running on the planned future CTA site in Chile for a year while two others are expected to become operational before the end of 2018. We report on the hardware and current status of operation and/or deployment of all the FRAM instruments in question as well as on some of the preliminary results of integral aerosol measurements by the FRAMs in Argentina and Chile.}

\maketitle

\section{Hardware and operational status}

\subsection {Auger FRAM}

The first FRAM (F/(Ph)otometric Robotic Atmospheric Monitor) telescope was installed in 2005 at the Pierre Auger Observatory in Malarg\"{u}e, Argentina \cite{AUGER} and has undergone several hardware iterations and explored several methods of atmospheric monitoring using stellar photometry \cite{Hindawi}. Since 2009 it is routinely used for ``rapid monitoring'', that is to determine the possible presence of clouds in the direction of a recently recorded ultra-high-energy cosmic-ray shower \cite{rapid,Malaga1}. The dataset thus generated has inspired the idea of the determination of the vertical aerosol optical depth (VAOD) using series of wide-field images at different elevations (``scans'') \cite{VAOD, EBR}. To this end, the Auger FRAM routinely takes such scans even in the absence of a cosmic-ray trigger since the beginning of 2016; however only a fraction of the scans taken can be used for aerosol determination so far because even scattered distant clouds (otherwise irrelevant for the Pierre Auger Observatory itself) in the distance interfere with the fitting procedure and those are a relatively common occurrence at the Auger site.

The current Auger FRAM setup \cite{Malaga2} has been installed in February 2013. The key part for the VAOD analysis is the ``wide-field camera'' which consists of a Moravian Instruments large-format G4-16000 CCD and a Nikkor 300/2.8 lens, together yielding a $7\,^{\circ}\times7\,^{\circ}$ field of view over 16 Mpix; the installation of this instrument marks the beginning of high-quality VAOD data. A major upgrade of the ``narrow-field camera'' is being carried out in September 2018: the replacement of the current 12-inch Meade SCT with a small G2-1600 CCD with the Orion UK 12-inch ``Optimised Dall-Kirkham'' telescope with ASA 3-inch focuser and another G4-16000 CCD will increase the field of view tenfold to roughly 1 sq. degree, possibly allowing VAOD studies with the large telescope (which has so far been used mainly for supplementary astronomical observations). During the upgrade, the Paramount ME mount will be retired after almost 10 years of service under heavy load and replaced with a new 10Microns GM2000 mount. While the Paramount worked reliably after a proper maintenance schedule was established, the 10Micron mount promises at least 10 years of maintenance-free operation, easier handling of the relatively heavy setup as well as absolute encoders on both axes allowing for additional level of safety in remote operation.

\begin{figure*}
\begin{centering}
\includegraphics[width=\textwidth]{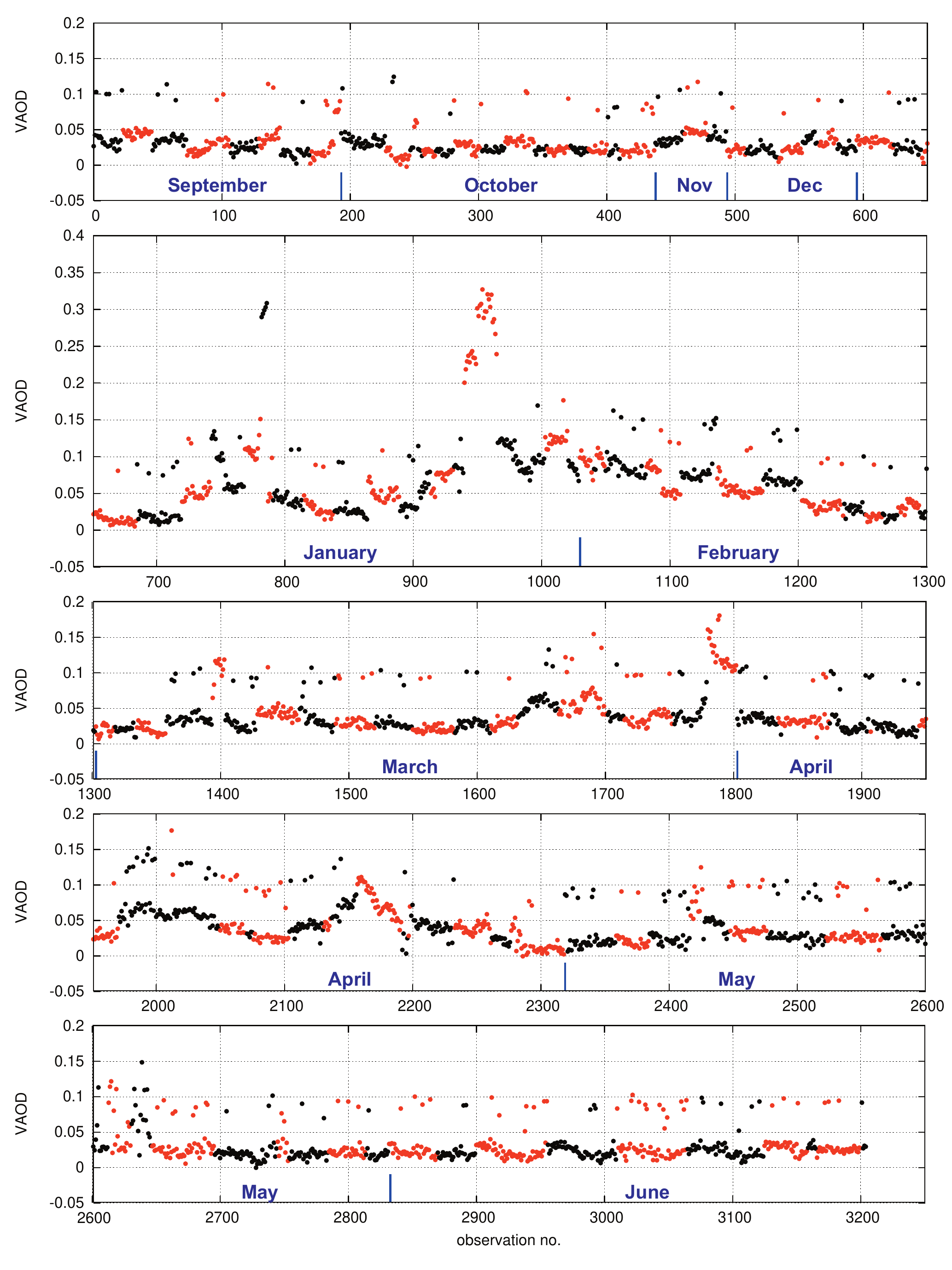}
\caption{Time series of VAOD measurements taken by the CTA FRAM telescope located near Cerro Paranal in Chile from 09/2017 to 06/2018. Each point represents one scan in altitude from the zenith to the horizon in seven 30s exposures. Consecutive points of the same color were taken during the same night, but there can be several nights skipped due to weather or hardware issues for each color change.}
\label{cta}
\end{centering}
\end{figure*}

\begin{figure*}
\begin{centering}
\includegraphics[width=\textwidth]{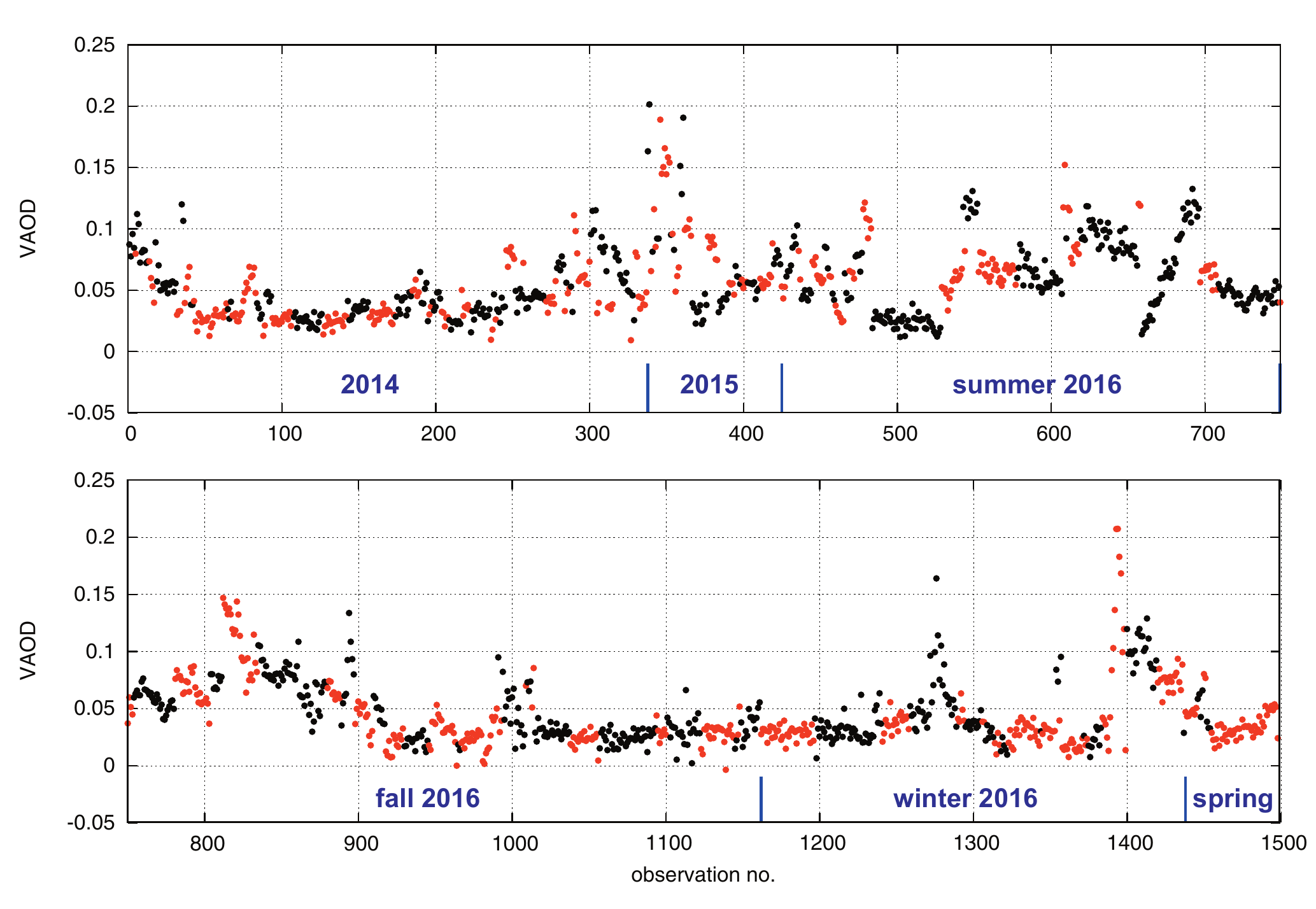}
\caption{Time series of VAOD measurements taken by the Auger FRAM telescope located near Malargue, Argentina from 03/2014 to 11/2016. Each points represents one scan in altitude from 30 degrees elevation to the horizon in at least five 30s exposures. Consecutive points of the same color were taken during the same night, but there can be many nights skipped due to weather or hardware issues for each color change.}
\label{auger}
\end{centering}
\end{figure*}

\subsection {CTA FRAMs}

The FRAM concept has been proposed to be integrated within the atmospheric monitoring for the future Cherenkov Telescope Array gamma-ray observatory \cite{CTA}. During the CTA operation, the FRAM would continuously monitor the field(s) of view of the Cherenkov telescopes for any sign of changes in atmospheric transmission. However, a number of altitude scans is also foreseen for nightly operation as those provide direct self-calibration for the FRAM \cite{CTAfram}. To get a complex view of the atmospheric properties of the future CTA sites, several devices are being deployed ahead of time, including the FRAMs: one is already deployed on the future CTA-South site near Cerro Paranal in Chile, the second one is almost completely assembled in Prague and will be probably installed on the future CTA-North site on La Palma, Canary Islands in October/November 2018; a third FRAM, again for the Chilean site, should follow within several months. When deployed ahead of the CTA itself, the FRAMs, in the absence of other duties, take altitude scans continuously, resulting in large amounts of data; as the Chilean site has exceptionally clear weather, the majority of those can be used for VAOD determination.

The wide-field detectors for all three CTA FRAMs are identical: again a G4-16000 CCD but with a Zeiss 135/2 lens, extending the field of view to $15\,^{\circ}\times15\,^{\circ}$. The two FRAMs in Chile use Paramount MYT mounts, which are probably the lightest robotic mounts on the market that satisfy requirements for remote use. The FRAM for La Palma uses 10Microns GM2000 as it will be accompanied by a narrow-field telescope previously used in the BART project of the Institute of Astronomy of the Czech Academy of Sciences; this telescope will be removed before the CTA operation commences. All three setups have been or will be extensively tested on the provisional platforms in the courtyard of the Institute of Physics in Prague and their components measured for spectral properties in the optical laboratory in Olomouc and for non-linearity of the CCDs in Prague.

\section{VAOD Data}

All the FRAM setups are equiped with BVRI photometric filters, but most of the measurements (and all of those presented in this section) are taken through the B filter. The B filter is the closest easily accessible standard photometric filter to the relevant wavelength range for both Auger and CTA as the fluorescence lines observed by Auger lie between 300 to 400 nm and this is also where the atmospheric Cherenkov spectrum peaks.The effective wavelength for VAOD measurements in B filter is between 430 and 440 nm depending on the type of aerosols and the specific hardware setup. Reaching even slightly shorter wavelenghts is difficult because it requires custom optics and detectors instead of off-the-shelf components and there is no directly applicable all-sky stellar catalog for that purpose.

Fig.~\ref{cta} shows all currently available VAOD measurements from the CTA FRAM. Apart from a period in October/November 2017 when the roof of the telescope housing was not working due to improper reaction of the firmware to low oil levels and some small downtime due to internet connection issues, the telescope was operated continuously without issues until the cooling of the camera stopped working in June 2018, thus producing a large amount of measurements, from which roughly a half has been processed so far due to limitations on computer power on site and bandwidth to transfer images. As the roof is a prototype of the unified housing system for all the CTA FRAMs, it is acceptable to show issues and the failure was used to develop better handling of unexpected conditions in the firmware; the camera was replaced in July 2018, but new data taken since then will have to be analyzed separately, because the necessary corrections are different for each individual camera. The cooling issue has been worked out with the vendor and all cameras liable to this problem are being upgraded.

The data series show an obvious population of outliers of unknown origin. So far we are not able to reject those based purely on a single observation, but the comparison with the Sun/Moon Photometer measurements (see \cite{EBR}) show that these are likely of instrumental origin. Barring the outliers, the conditions on the future CTA site are often stable for many days with very small aerosol content.

Fig.~\ref{auger} shows selected VAOD measurements from the Auger FRAM. The Auger FRAM underwent several hardware changes, including several changes of the camera due to vendor improvements in the environmental protection of the camera electronics and reliability of the filter wheel. For each camera, the corrections for non-linearity and bias signal described in \cite{EBR} have to be developed separately and for now, only the data from the period 03/2014 to 11/2016 have been fully processed. In the first two years, the observations were only carried out for rapid monitoring purposes and thus they depend on the triggers from the Observatory; due to problems with the trigger, the 2015 data are limited. Moreover, most of the triggers are designed to look for anomalous showers, which can be often caused by clouds -- this means that the rapid monitoring triggers specifically find clouds on the sky, further limiting the amount of data available for aerosol analysis. Since the implementation of dedicated aerosol observations in early 2016, the amount of data available increased significantly and yet even more data from 2017 and 2018 are awaiting processing with the non-linearity corrections. The aerosol content of the Auger site is more variable, particularly in austral Summer, when thunderstorms and other severe weather may occur, however significant stable periods also exists on the site. Interestingly, the outlier problem found on the CTA FRAM does not seem to occur at the Auger FRAM.
\\
\section*{Acknowledgments}
The authors are grateful for the support by the grants of the Ministry of Education, Youth and Sports of the Czech Republic, projects LTT17006, LM2015046, LTT18004, LM2015038, CZ.02.1.01/0.0/0.0/16\_013/0001402, CZ. 02.1.01/0.0/0.0/16\_013/0001403 and CZ.02.1.01/0.0/0.0 /15\_003/0000437. This research was supported in part by the Polish National Science Centre under grant 2017/26/D/ST9/00449. The work is partially performed according to the Russian Government Program of Competitive Growth of Kazan Federal University.

\end{document}